\documentstyle[prl,aps,multicol]{revtex} 
\begin{document}
\draft

\title{Vortex lattice melting, phase slips and decoupling
 studied by microwave dissipation in
$\bf Bi_{2}Sr_{2}CaCu_{2}O_{8+\delta}$ }

\author{H.\ Enriquez, Y.\ De\ Wilde, N.\ Bontemps}
\address{Laboratoire de Physique de la Mati\`ere Condens\'ee, Ecole Normale
Sup\'erieure, 24 rue Lhomond, 75231 Paris Cedex 05, France}

\author{T.\ Tamegai}
\address{Department of Applied Physics, The University of Tokyo,
Hongo, Bunkyo-ku, Tokyo 113-8656, Japan}


\maketitle
\begin{abstract}
We have measured the microwave dissipation at 10~GHz in a 
$\rm Bi_{2}Sr_{2}CaCu_ {2}O_{8+\delta}$ single crystal, between 75 and 50~K,
as a function of the static magnetic field applied parallel to the c axis.
We observe a clear-cut onset in  the dissipation precisely at the melting
field. In the liquid phase, for fields $\sim$ 1 -2 kG, the in-plane and the
c-axis resistivities deduced from microwave absorption display a different
temperature behavior, in contrast with DC measurements. Our c-axis high
frequency data and the DC results can be reconciled by including phase slips
 due to the motion of pancake vortices,
as suggested by Koshelev. The  microscopic origin of the c-axis
dissipation is thus identified through its frequency dependence. 
\end{abstract}

\begin{multicols}{2}
\narrowtext
In high temperature superconductors, thermal fluctuations play a crucial
role in the behavior of the vortex system, inducing vortex-lattice
phase transitions and new phases in the (H,T) phase diagram of the mixed
state \cite{nelson}.
A first-order transition corresponding to melting of the vortex lattice
has been evidenced through a discontinuity in the entropy \cite{schilling},
at the same
field $H_m$ where a step appears in the static magnetization, in  clean
$\rm Bi_2Sr_2CaCu_2O_{8+\delta}$ (BSCCO) \cite{pastoriza1} 
and $\rm YBa_2Cu_3O_{7-\delta}$ (YBCO) single crystals\cite{welp}.
Local measurements using miniature Hall probes locate
very precisely this melting transition in BSCCO single crystals
down to 40K \cite{zeldov}. DC measurements show that the in-plane resistance
drops down to zero, within the experimental resolution,  in YBCO
untwinned crystals and in BSCCO at the melting transition
\cite{kwok,charal,fuchs1,welp}. The c-axis DC resistance  exhibits a
similar sharp decrease at the melting temperature $T_m$
\cite{kadowaki,fuchs2}.  
It has been suggested that the latter
force free ($H \parallel j_{c} \parallel$ \^{c})
dissipation bears some  analogy with the voltage induced in a
single Josephson junction by thermally activated phase slips occuring
near $T_c$
\cite{ambegaokar,deutscher,glazman,daemen,briceno,gray,hellerquist,cho,kadowaki}, inasmuch as BSCCO can be described as a stack of
small area Josephson junctions \cite{kleiner}.
 Koshelev recently suggested that in-plane diffusion of pancake
vortices \cite{koshelev} is the mechanism which accounts for the similar
temperature dependence of the ab-plane and c-axis DC conductivities measured
above $H_m$  \cite{latyshev,busch,livanov}.
In-plane diffusion induces phase slips
between adjacent layers, leading to a loss of phase coherence
and thus
giving rise to the so-called decoupling transition.
It is presently still debated, from an experimental
\cite{pastoriza1,kadowaki,fuchs2,hellerquist,cho,doyle}
and theoretical point of view \cite{daemen,glazman,blatter}, 
 whether decoupling occurs at the  melting field or above.
In this controversy, the c-axis dissipation mechanism plays
a key role, as far as it may probe the phase coherence between the layers. 

In this paper, we report 10 GHz microwave dissipation measurements
carried out in a BSCCO single crystal in the vicinity of the
melting line, which is clearly
identified through static magnetization measurements. 
 The unique  features of our experiment are:
 i) it locates precisely $H_m$ by the simultaneous onset of
in-plane and out-of-plane microwave dissipation.
 ii) a quantitative estimate of both contributions shows that
  the ab plane microwave absorption exhibits a
thermally activated behavior as found in DC measurements,
whereas the c-axis microwave absorption, unlike the DC resistivity
exhibits a different temperature behavior.
This points to a frequency dependent
resistivity associated with diffusion induced phase slips, as
suggested in ref.\cite{koshelev}.

Both magnetization and microwave dissipation measurements were performed in
the same BSCCO single crystal, shaped into a rectangular
platelet ($\rm {2 \times 0.8 \times 0.03}\,mm^3$),
 with $T_c$=84K and a transition width
(measured at 10 GHz) $\Delta T_c \sim 0.4 \,{\rm K}$.
Figure 1 shows the result of dc magnetization measurements
performed with a Quantum Design SQUID magnetometer. The inset displays 
a typical zero field cooled and field cooled magnetization curve
at 65K as a function of magnetic field. The data exhibit a "step"
in the reversible regime, which allows to locate $H_m$ 
by a lower and upper value, as shown in the inset of fig.1.
The resulting melting line is shown in fig.1 between 75 K
and 45K. The Clausius-Clapeyron equation for a magnetic first order
transition yields the entropy change
$\Delta S \sim 1 \,{\rm k_B}$ per vortex and per $CuO_2$ layer,
consistent with other reported values \cite{zeldov}.

The experimental set-up used to measure the microwave dissipation as
a function of field (0-2 kG) and temperature (45-75 K) has been described
elsewhere \cite{enriquez}. The geometry is depicted
in fig.2. The sample is placed in a $\rm TE_{102}$ resonant
cavity (${\omega \over{2 \pi}} = 9.6\,\rm GHz$) of an Electron Spin Resonance (ESR)
ESP 300E Br\"{u}ker spectrometer, at the
maximum of the microwave magnetic field $h_1(\omega)$ and in zero electric field.
The static field H is applied parallel to the c axis.
As $h_1(\omega)$ is parallel to the ab plane (fig.2), microwave currents
flow within the ab plane and along the c axis. They generate a
magnetic moment $m(\omega)$, and hence a macroscopic susceptibility
$\chi(\omega)=m(\omega)/h_1(\omega)$. Following Landau \cite{landau},
in the ohmic regime $j(\omega)= {1 \over \rho(\omega)}E(\omega)$,
(we have checked that when changing the magnitude of $h_1$
 from
5 mOe to 50 mOe, the response is indeed ohmic within our experimental
resolution), the energy W dissipated by the sample writes:

\begin{eqnarray}
{\rm W}={1 \over 2} {\mu_0\, \omega\, h_1^2 \,\chi ''(\omega)\, V}
\label{eq1}
\end{eqnarray}

$\chi ''(\omega)$ is the imaginary part of the susceptibility. $\it {V}$ is
the volume of the sample. The measured quantity is
$\delta \chi '' = \chi '' \rm (H) - \chi '' (H=0)$.
As explained further, $\chi '' \rm (H=0)$ is negligible compared to
$\chi '' \rm (H)$, hence  $\delta \chi '' =\chi '' \rm (H)$.  
The  absolute value (dimensionless in SI) of $\chi ''$ has been calibrated with
respect to a PbIn(6\%) platelet (same geometry as the BSCCO crystal), within
$20\,\%$ accuracy.

It is easy to show that as long as the depths over which the ab and c currents penetrate
are small with respect to the dimensions along the same direction 
("thick limit" approximation), W can be rewritten as:

\begin{eqnarray}
{W }
\approx { h_{1}^2}{ (  R_{s,ab} d_{1} d_{2} +  R_{s,c} d_{i}e )}
\label{eq2}
\end{eqnarray}

with i=1(2) refering to position 1(2), as defined in fig.2.
$d_{1} =2\, \rm mm$, $d_{2} =0.8\, \rm mm$.
$R_{s,k}$ is the surface resistance probed by the current flowing along the
k direction (fig.2). Two measurements corresponding to two
geometries of the sample are necessary
in order to estimate $R_{s,ab} $ and $R_{s,c}$. Here,  we simply rotate
the sample by
$90^{\circ}$ around the c axis:  e (crystal thickness) is unchanged,
whereas d (platelet size) changes by a factor 2.5. It is clear from eq.2
that position 1 gives more weight to dissipation due to c-axis currents.

Figure 3 shows the microwave dissipation measured as a function of field
at $75\,K$ and $50\,K$ in position 1 and 2.
At a field  $\rm H^{\ast}$ indicated by the arrow, we observe a
clear change in the dissipation regime: the dissipation is essentially
zero within our experimental resolution ($\sim 1.10^{-3}$) up to this
field, then it starts to increase steadily
reaching $\sim 10^{-2}$ at $400\,G$; this figure represents
$\sim 10^{-1} \chi ''_n$ where $\chi ''_n$ is the susceptibility in the
 normal state ($T \geq T_c$).
Systematic measurements as a function of H have been
performed from $75\,K$ down to $50\,K$ and display a similar change at
fields $\rm H^{\ast} (T)$. These fields are reported in fig.1:
the line  $\rm H^{\ast} (T)$  coincides with the melting line
determined by magnetization data. This establishes experimentally
that the thermodynamic first-order transition in the vortex
lattice induces a clear-cut onset in  microwave dissipation.
The next important result is that the dissipation at  $H > \rm H^{\ast} (T)$
is larger in position 1 than in position 2, which
suggests a systematic contribution of c-axis currents.
One could think of this absorption being due to
Josephson plasma resonance (JPR) \cite{tsui,matsuda}. Josephson plasma
modes were carefully studied in the same configuration ($h_1$ parallel to
the ab plane) at 35 GHz microwave frequency \cite{kakeya}: a sharp resonance
in the field range 1 to 7 kOe (depending on temperature) , is assigned to
the longitudinal mode, whereas a much broader absorption is observed
at much higher fields (up to 6 T). Since JPR frequency is expected to vary
 as $H^{-1/2}$, for 10 GHz such features  would appear at
fields roughly ten times larger, e.g. entirely out of our experimental
field range. Two other remarks confirm that we can rule out JPR as the
absorption process in our experiment.
i)the low tail of
such an absorption should already  be visible in the solid phase,
in contrast with the onset of dissipation occuring unambiguously at
the melting field (this point has been checked in 5 other crystals),
ii) we have swept the field up to 1T and we observe a steadily increasing,
slowly saturating  signal, with no hint of an absorption peak \cite{matsuda}.
Therefore we think that we can rule out this absorption mechanism. We then
rely on eq.2 in order to perform  quantitative estimates.
Checking the "thick limit" approximation requires the evaluation of the surface
resistance. The starting point is the surface impedance, which reads
\cite{coffey}:

\begin{eqnarray}
{Z_s}={R_{s}+iX_{s}}={i \omega \mu_{0} \tilde{\lambda}}
\label{eq3}
\end{eqnarray}

where $\tilde{\lambda}$ is the complex penetration depth.  $\tilde{\lambda}$
depends on the London penetration depth, the normal fluid skin depth,
and the complex effective skin depth arising from vortex motion \cite{coffey}.

 Surface impedance  measurements show that the dissipation in the London
regime is $\chi ''_L \sim 10^{-4}$, i.e. far below
our experimental resolution. Therefore, we shall neglect the normal fluid
contribution. The microwave dissipation due to vortices above the melting field,
whether it arises from a flux flow mechanism \cite{gittleman} or
from a thermally activated process, as we will see,
\cite{latyshev,busch,livanov,palstra}, can be described by an effective
resistivity $\rho_v$, (which needs not be the dc resistivity). We finally
neglect the contribution of the Campbell penetration depth, since the magnetization is
reversible above $H_m$.
Therefore the general expression given in ref. \cite{coffey}
for $\tilde{\lambda}$ reduces to:

\begin{eqnarray}
{\tilde{\lambda}^2} \simeq \lambda^{2} -{i \over 2} \delta_{v}^2
\label{eq4}
\end{eqnarray}

$\lambda$ is the London penetration depth, and $\delta_v$ is the vortex induced
skin depth defined as usual  by
${\delta_v} = ( { {2 \rho_v} / {\mu_0 \omega} } ) ^{1\over2}$.
 The London penetration depths 
$\lambda_{ab}(T)$ and $\lambda_{c}(T)$ for BSCCO  at $T=0 K$ are not
 precisely  known.  Quoted values of $\lambda_{ab}(0)$ vary
from $1800 \AA$ up to 3000 \AA \cite{jacobs}. $\lambda_{c}(0)$ is still more
uncertain. We have taken the value derived from our own measurements of
the first penetration field of Josephson vortices, namely
$\lambda_{c}(0) = 220 \mu m$ \cite{enriquez3}.
The temperature variation
$\Delta \lambda_{ab}(T)$ and $\Delta \lambda_{c}(T)$ have been  established
by a cavity perturbation technique \cite{jacobs}. In the least favourable case,
$\lambda_{c}(75K) = 280 \mu m < d_{2}=800 \mu m$, and
 $\lambda_{ab}(75K) = 0.6 \mu m \ll e=20 \mu m$.
From the estimates of the resistivity at 2 kG described below, we find
that
$\delta_{v,ab} \sim 2\mu m$, and $\delta_{v,c} \sim 80 \mu m $, hence again small
with respect to $e$ and $d_2$ respectively. Therefore the "thick limit" 
 approximation is appropriate.

Combining eq.2, 3 and 4, we have computed $\rho_{ab}\rm (H=2\,kOe)$ and
$\rho_c\rm (H=2\,kOe)$ (we chose this  field value
in order to achieve a better accuracy; we checked that the results are similar for
$\rm H=1\,kOe)$.
We have reported  $\rho_{ab}$ and $\rho_c$  in fig.4 as a function of
$\rm 1/T$. Changing $\lambda_{ab}(0)$ hardly changes  $\rho_{ab}$
($\leq 10 \%$).
$\rho_c$ has been deduced using two very different values for
$\lambda_{c}(0)$, without changing the main trends:
i) $\rho_{ab}$ exhibits a thermally activated
behavior which compares very satifactorily to dc resistivity
measurements  \cite{palstra,busch,fuchs1}. This suggests that
 10 GHz lies below the depinning frequency
\cite{gittleman} of pancake vortices in BSCCO (which to the best of our
knowledge, has not been determined yet).
ii) $\rho_c\rm (H=2\,kOe)$ does not display
the same temperature dependence as   $\rho_{ab}$, in contrast
with dc resistivity data \cite{busch,kadowaki}. 
Koshelev suggested that the c-axis conductivity at frequency  $\omega$
may be related to the dc conductivity through the characteristic phase slip
frequency. The resistivity at finite frequency reads then \cite{koshelev}:
 
\begin{eqnarray}
{\rho_{c}(\omega=0)}= {\rho_{c}(\omega)}{\omega_{ps}^2 \over {\omega^2 +\omega_{ps}^2 }}
\label{eq5}
\end{eqnarray}

The phase slip frequency value  was given in \cite{koshelev}:
$\omega_{ps} \simeq 2\times 10^{8} T \rho_{ab} ln(nr_{max}^2)$, where T is in K,
$\rho_{ab}$ in $\mu \Omega.cm$, n is the density of mobile pancakes, and
$r_{max}$
is the distance over which the phase within a layer is no longer sensitive to
pancake motion. Using our own $\rho_{ab}$ values for
consistency, we have computed $\rho_{c}(\omega=0)$ from eq.5, adjusting
$r_{max}$ so as to retrieve a similar temperature dependence for
 $\rho_{c}(\omega=0)$
and $\rho_{ab}$.  The result is shown in fig.4: within our experimental accuracy,
 the temperature dependence of
$\rho_c$ is now consistent with the dc data.
The above procedure yields  $ln(nr_{max}^2) \sim 8$, hence $r_{max} \sim 50$,
in units of inter vortex distance at $\sim 2 kOe$, if all vortices are mobile.
 Although this figure is
difficult to estimate, the value we find seems reasonable.

The phase slips which yield the c-axis dissipation may also be responsible for
the decoupling process. 
It is worthwhile discussing the question of decoupling in this context:
Daemen defines the decoupling field as the field where the supercurrent
along the c axis vanishes \cite{daemen}.  However, this
theoretical description relies on the estimate of the phase difference
between adjacent layers. Therefore the critical current which is calculated
self-consistently is a local quantity. Indeed, Daemen specifies that at the
decoupling field, vortex lines  dissociate into a gas of pancakes. In such
a case, as already suggested \cite{hellerquist}, one expects that the
normal state resistance is recovered. This occurs in our experiment
at much higher field: 2000G at 75K or 3000G at 70K. Therefore there is an
essential difference between the onset of decoupling (which takes place
at the melting field) where phase slips induce dissipation but
phase coherence is not lost between adjacent layers \cite{glazman}, and
full decoupling.

In conclusion, we have identified  within a single experiment
the first order melting transition, the c-axis resistivity and the decoupling
phenomenon. Our analysis which accounts for the effect of frequency,  is
consistent with a recent theoretical suggestion which assigns the same microscopic
origin to in-plane and c-axis dissipation, namely thermally assisted phase slips.
The picture which appears to be confirmed experimentally is that
phase slips are induced by the thermally assisted diffusive motion of pancake
vortices: the same mechanism might be responsible for melting, for dissipation
along the c axis and for decoupling.\\

We are very much indebted to  A.~Koshelev for his illuminating remarks
dealing with the evaluation of our data.
We thank P.~Monod for his numerous suggestions in the course of this work.
We gratefully acknowledge the support of K.~Behnia. 
This work was partly supported by the Grant-in-Aid for Scientific Research
from the Ministry of Education, Science, Sports and Culture, Japan.
The laboratory Physique de la Mati\`ere Condens\'ee is associated to
CNRS and Universities Paris~VI and Paris~VII. \\
\begin{figure}
\caption{First order melting line defined from dc magnetization
measurements and from the onset of microwave dissipation. The arrows in
the inset locate the field where the the jump of the dc
magnetization starts ($\rm H_{m,low}$) and ends ($\rm H_{m,up}$).
\label{melting}}
\end{figure}
\begin{figure}
\caption{Sketch of the experimental geometry. The microwave field $h_1$
lies parallel to the ab plane. The static field H is parallel to the c
axis. Microwave currents ($j_{ab}$ and $j_c$) flow  is displayed
(seen from above).
\label{geometry}}
\end{figure}

\begin{figure}
\caption{Dissipation as a function of the applied field at 75K (a)
and 50K (b) in position 1 and 2. The onset of dissipation, indicated by
the arrow, occurs at the melting field. The ratio
$\chi ''_1 / \chi ''_2$ increases as the temperature decreases,
reaching 2.4 at 50K, which means that the dissipation due to ab
currents has become negligible (see text).
\label{chi}}
\end{figure}
\begin{figure}
\caption{Resistivities $\rho_{ab}(\omega, \rm H=2\,kOe)$ (full circles) and
$\rho_{c} (\omega, \rm H=2\,kOe)$ (full triangles) versus $1 /T$.
The solid line shows the Arrhenius behavior of $\rho_{ab}$.
Values used for the London penetration depths are
$\lambda_{ab}(0)=2100 \AA$ (see text),
$\lambda_{c}(0)=220 \mu m$ \protect\cite{enriquez3} (down triangles) and
$40 \mu m$ \protect\cite{jacobs} ( up triangles).
Open triangles display c-axis resistivity data corrected for
the effect of frequency (see text).
\label{rho}}
\end{figure}
\end{multicols}

\begin{references}
\bibitem{footnote}    At very low fields (50K), the small peak  is the
signature of the lock-in transition, studied elsewhere \cite{enriquez},
which is induced here by a slight misalignment of the static field with
respect to the c-axis ($\sim 1^{\circ}$). For the present study, this peak is
not relevant.
\end{references}
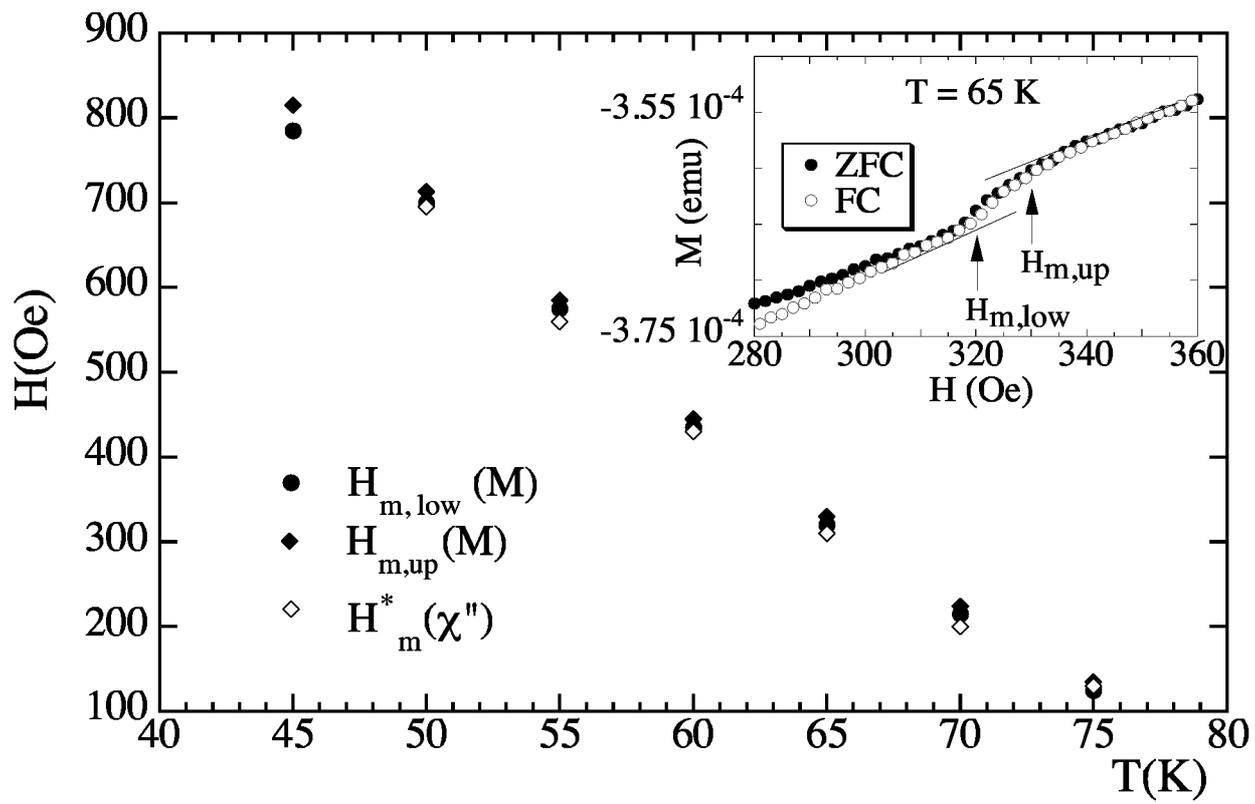
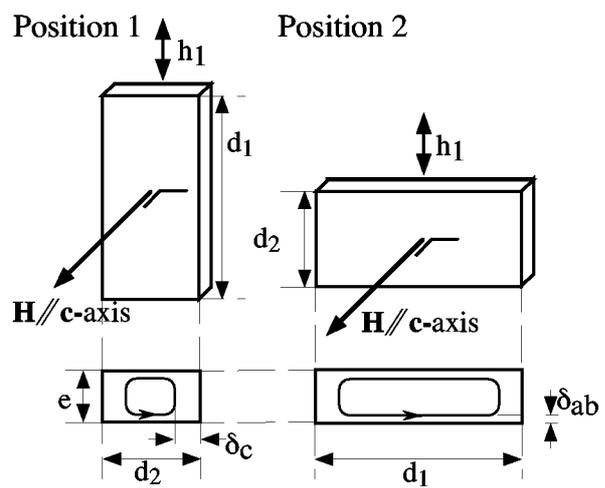
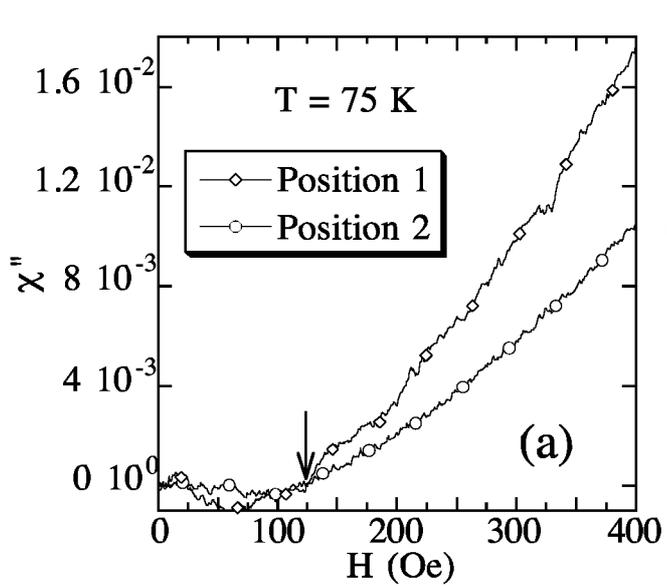
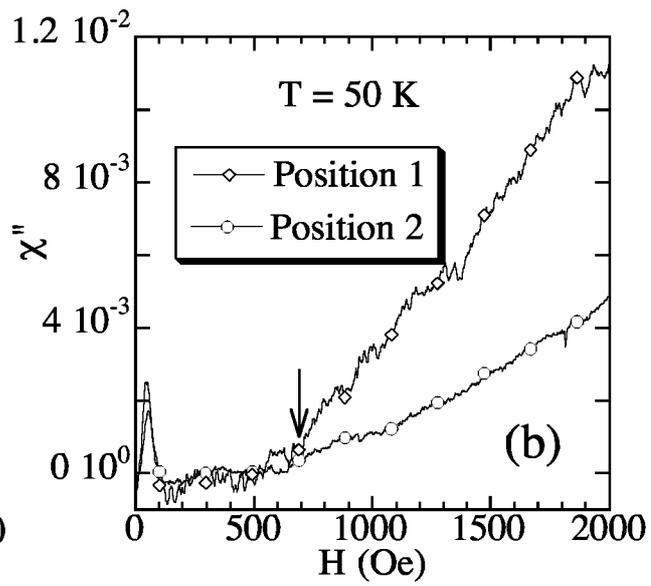
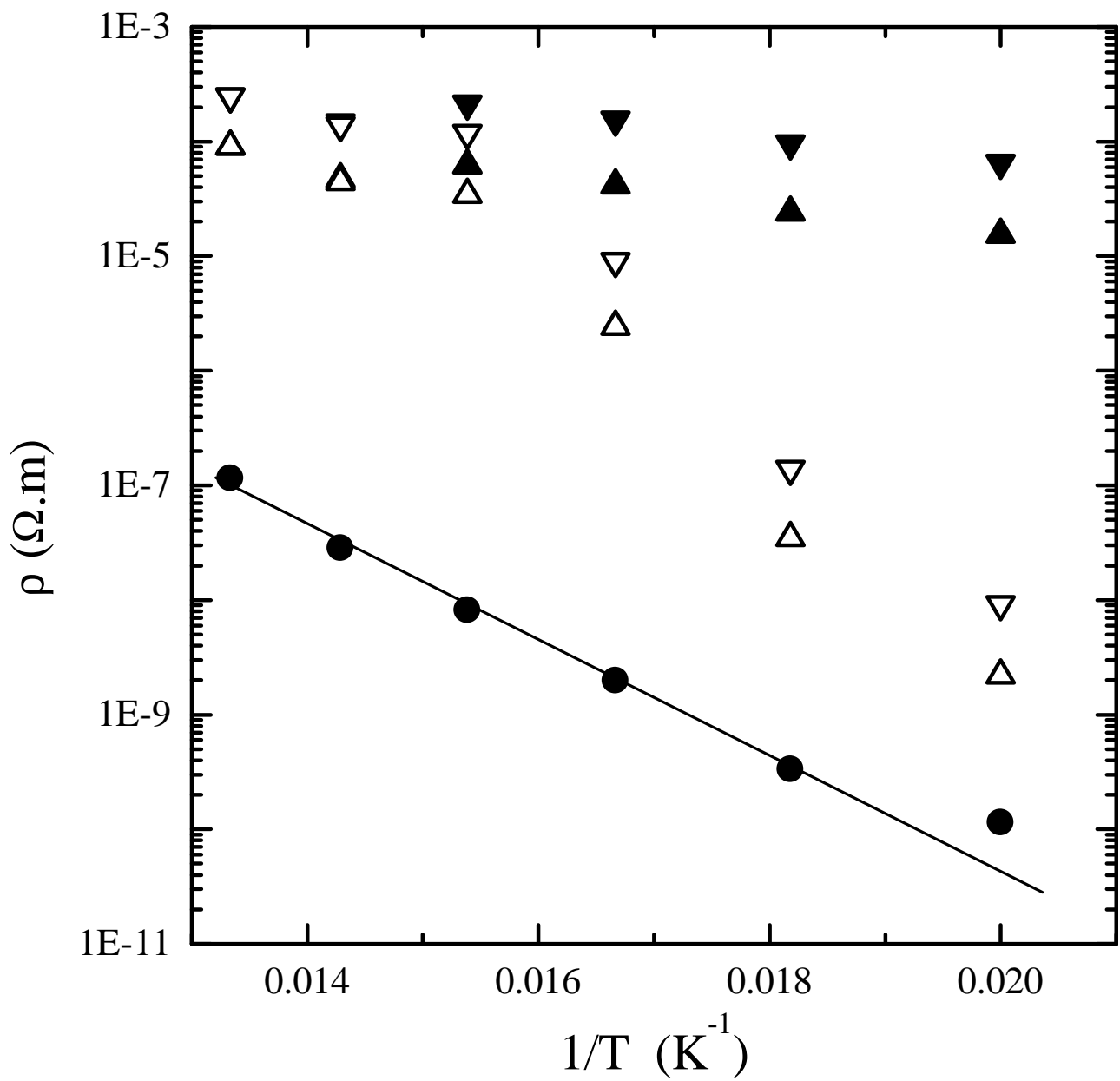
\end{document}